\documentclass[preprint,12pt]{elsarticle}
\journal{Mathematics and Mechanics of Solids}
\usepackage{graphics,epsfig}
\usepackage{graphicx}
\usepackage{float}
\usepackage{amssymb,amstext,amsmath}
\usepackage{mathtools}
\usepackage{booktabs}
\usepackage{physics}
\begin{document}
\begin{frontmatter}
\title{A lower bound for the energy decay rate in piezoelectricity}
\author{K. C. Le$^{a,b}$\footnote{Email: lekhanhchau@tdtu.edu.vn}}
\address{$^a$Division of Computational Mechanics, Institute for Advanced Study in Technology, Ton Duc Thang University, Ho Chi Minh City, Vietnam\\
$^b$Faculty of Civil Engineering, Ton Duc Thang University, Ho Chi Minh City, Vietnam}
\begin{abstract} 
This paper establishes a lower bound for the energy decay rate in piezoelectric cylinders. The bound incorporates material properties and geometric factors, including the cross-section's Poincaré-Wirtinger and Korn constants. A detailed analysis of a circular cross-section cylinder yields a precise numerical lower bound, illustrating the practical application of this result.
\end{abstract}


\end{frontmatter}

\section{Introduction}
Saint-Venant's principle, dating back to 1856 \cite{de1856memoire}, provides a fundamental insight into the decay of stress in an elastic bar subjected to self-balanced loads at its ends. Toupin \cite{toupin1965saint} established the first rigorous proof of this principle, demonstrating the exponential decay of stress and energy with distance from the loaded edge and relating the decay rate to the cylinder's smallest natural frequency of vibration. This foundational work was later extended to piezoelectricity by Batra and Yang \cite{batra1995saint}, where the energy decay rate similarly depends on the smallest natural frequency. However, determining this frequency can be challenging for arbitrary cross-sections, and counterexamples exist where the decay rate becomes arbitrarily small, highlighting the need for meaningful lower bounds.

Berdichevsky \cite{berdichevskii1974proof} addressed this challenge by deriving a lower bound for the stress and energy decay in a circular cylinder, demonstrating substantial decay over distances comparable to the cross-section radius. This paper builds upon Berdichevsky's work, extending his results to the realm of piezoelectricity. We first establish a novel variational formulation for the equilibrium of a piezoelectric body under self-balanced traction, employing a Dirichlet boundary condition for the electric vector potential to facilitate the derivation of a lower bound. Then, by leveraging this formulation, along with lower bounds for the internal and complementary energy densities and an energy identity for sub-bodies, we derive a lower bound for the energy decay rate in a piezoelectric cylinder. This bound incorporates the Poincaré-Wirtinger and Korn's second inequalities, combined with Berdichevsky's inequalities \cite{berdichevskii1973energy,berdichevskii1974proof}, and is explicitly calculated for cylinders of circular cross-section.

\section{Variational principle for piezoelectric body under mechanical loading}

We consider an inhomogeneous piezoelectric body occupying a three-dimensional domain $\mathcal{V}$ bounded by a piecewise smooth boundary $\partial \mathcal{V}$. The body is assumed to be in equilibrium under the action of mechanical loads, with no body forces present. The kinematic and equilibrium equations of the piezoelectric body as a continuous medium read \cite{landau2013electrodynamics}:
\begin{equation*}
\begin{split}
&\div \vb*{\vb*{\sigma }}=\vb{0},\quad \vb*{\varepsilon }=\grad_\text{sym} \vb{w}\equiv \frac{1}{2}( \grad \vb{w}+(\grad \vb{w})^T),
\\
&\curl \vb{E}=\vb{0},\quad \div \vb{D}=0.
\end{split}
\end{equation*}
Here, $\boldsymbol{\sigma}$ is the stress tensor, $\boldsymbol{\varepsilon}$ the strain tensor, $\vb{w}$ the displacement vector, $\vb{E}$ the electric field,  $\vb{D}$ the electric displacement (induction) field, while $\grad $ denotes the nabla operator and $(\grad \vb{w})^T$ is the transpose of $\grad \vb{w}$.

To describe the electromechanical behavior of the inhomogeneous piezoelectric material, we adopt a specific form of the internal energy density $U$:
\begin{equation*}
U(\vb{x},\vb*{\varepsilon},\vb{D})= \frac{1}{2} [\vb*{\varepsilon} \mathbf{:} \vb{c}^D(\vb{x}) \mathbf{:} \vb*{\varepsilon}- 2\vb{D}\vdot \vb{h}(\vb{x}) \mathbf{:} \vb*{\varepsilon}+\vb{D}\vdot \vb*{\beta }^S(\vb{x}) \vdot \vb{D}],
\end{equation*}
where $\vb{c}^D(\vb{x})$ is the fourth-rank tensor of elastic moduli at constant electric displacement, $\vb{h}(\vb{x})$ the third-rank tensor of piezoelectric moduli, and $\vb*{\beta}^S(\vb{x})$ the second-rank tensor of impermittivity moduli at constant strain. Note that $U(\vb{x},\boldsymbol{\varepsilon},\vb{D})$ is a strictly positive definite quadratic form of $\vb*{\varepsilon}$ and $\vb{D}$, ensuring that the energy is always non-negative. The constitutive equations, derived from the internal energy density, relate the stress and electric field to the strain and electric displacement:
\begin{equation}\label{constitutive}
\begin{split}
&\vb*{\sigma } = \pdv{U}{\vb*{\varepsilon }}=\vb{c}^D(\vb{x})
\vb{:}\vb*{\varepsilon } -\vb{D}\vdot \vb{h}(\vb{x}),
\\
&\vb{E} = \pdv{U}{\vb{D}}=-\vb{h}(\vb{x})\vb{:}
\vb*{\varepsilon } +\vb*{\beta }^S(\vb{x}) \vdot \vb{D}.
\end{split}
\end{equation}

We now focus on the case of purely mechanical loading, where the boundary $\partial \mathcal{V}$ is subject to prescribed traction. Specifically, traction $\vb*{\tau}$ is applied on a portion $\mathcal{S}_\tau$ of the boundary, while the remaining part $\mathcal{S}_f$ is traction-free. Since there is no electrode at the boundary, we pose the condition of free surface charge
\begin{equation}
\label{Dbc}
\vb{D}\vdot \vb{n}=0 \quad \text{at $\partial \mathcal{V}$},
\end{equation}
where $\vb{n}$ is the outward unit vector normal to $\partial \mathcal{V}$. To ensure a well-posed problem and exclude rigid body motions, we impose the following conditions and constraints:
\begin{align}
\label{sequil}
&\int_{\mathcal{S}_\tau} \vb*{\tau}\dd{a}=\vb{0},\quad \int_{\mathcal{S}_\tau} \vb{x}\cp \vb*{\tau}\dd{a}=\vb{0},
\\
\label{displ}
&\int_{\mathcal{V}} \vb{w}\dd[3]{x}=\vb{0},\quad \int_{\mathcal{V}} \curl \vb{w}\dd[3]{x}=\vb{0}.
\end{align}
Conditions \eqref{sequil} mean that the traction $\vb*{\tau}$ is self-balanced, while constraints \eqref{displ} guarantee the vanishing rigid-body translations and rotations. Using the results of Borchers and Sohr \cite{borchers1990equations}, we introduce a vector potential $\vb*{\psi}$ to represent the electric displacement as $\vb{D}=\curl \vb*{\psi}$, with 
\begin{equation}
\label{constraints1}
\div \vb*{\psi}=0,\quad \vb*{\psi}|_{\partial \mathcal{V}}=\vb{0}.
\end{equation}
The Dirichlet conditions \eqref{constraints1}$_2$ implies \eqref{Dbc}. This allows us to formulate the problem in terms of a variational principle (cf. \cite{batra1995saint,le1999vibrations}): the true displacement $\check{\vb{w}}$ and vector potential $\check{\vb*{\psi}}$ minimize the energy functional
\begin{equation}
\label{energyf1}
I[\vb{w}(\vb{x}),\vb*{\psi}(\vb{x})]=\int_{\mathcal{V}} U(\vb{x},\grad_\text{sym} \vb{w},\curl \vb*{\psi})\dd[3]{x}-\int_{\mathcal{S}_\tau} \vb*{\tau}\vdot \vb{w}\dd{a},
\end{equation}
subject to the constraints \eqref{displ} and \eqref{constraints1}. 

The existence and uniqueness of the minimizer of \eqref{energyf1} can be established by demonstrating that the functional \eqref{energyf1} is bounded from below, convex, and lower semi-continuous. While the convexity and lower semi-continuity of \eqref{energyf1} can easily be established, the boundedness from below involves utilizing the trace inequality \cite{adams2003sobolev}, Poincare-Wirtinger's inequality \cite{adams2003sobolev}, and second Korn's inequality \cite{horgan1995korn}:
\begin{equation}
\label{Korn-Poinc}
\begin{split}
& \int_{\partial \mathcal{V}} u^2\dd{a}\le C \int_{\mathcal{V}} (\grad u \vdot \grad u +u^2)\dd[3]{x} ,
\\
&\int_\mathcal{V} u^2\dd[3]{x} \le P \int_\mathcal{V} \grad u\vdot \grad u \dd[3]{x}
\\
&\int_{\mathcal{V}} (\grad \vb{w})\vb{:}(\grad \vb{w})\dd[3]{x}\le K \int_{\mathcal{V}} (\grad_\text{sym} \vb{w})\vb{:} (\grad_\text{sym} \vb{w})\dd[3]{x},
\end{split}
\end{equation}
where $C$, $P$, and $K$ are positive constants. Poincare-Wirtinger's inequality holds true for all scalar functions having zero average over $\mathcal{V}$, while second Korn's inequality holds true for all vector fields satisfying constraint \eqref{displ}$_2$. The existence and uniqueness of the minimizer follows a similar line of reasoning as in classical elasticity theory (see, for example, \cite{berdichevsky2009variational}).

\section{Energy decay in a functionally graded piezoelectric cylinder}

We consider a cylinder composed of a functionally graded piezoelectric material with materially uniform cross-sections. To facilitate the analysis, we adopt an index notation.  Let $\{x_1,x_2,x_3\equiv x\}$ denote a three-dimensional (3-D) Cartesian coordinate system, where $x$ is directed along the cylinder axis and the left edge of the cylinder lies in the $(x_1,x_2)$-plane. Greek indices (ranging from 1 to 2) represent components in the transverse plane, while Latin indices (taking values 1, 2, 3) denote components in the full 3-D Cartesian system.  Summation over repeated indices is implied, parentheses enclosing indices indicate symmetrization, and a comma preceding an index denotes partial differentiation. For brevity, we omit the index 3 in writing vector and tensor components, so that $w_3=w$, $\varepsilon _{33}=\varepsilon$, $\varepsilon_{3\alpha}=\varepsilon_\alpha$, $\sigma_{33}=\sigma$, $\sigma_{3\alpha}=\sigma_\alpha$, $D_3=D$ ect.

The internal energy density $U$ of the functionally graded piezoelectric cylinder with materially uniform cross-sections depends on the transverse coordinates $x_\alpha$ and the strain and electric displacement components:
\begin{equation*}
U=U(x_\alpha ,\varepsilon_{ij},D_i),
\end{equation*} 
where $\varepsilon_{ij}=w_{(i,j)}$ is the strain tensor and $D_i=\epsilon_{ijk}\psi_{k,j}$ is the electric displacement field, expressed in terms of the displacement $w_i$ and vector potential $\psi_i$, respectively, with $\epsilon_{ijk}$ the permutation symbol.

The cylinder is loaded by a self-balaced traction $\tau_i(x_\alpha,0)$ applied at its left edge $x=0$, with the remaining boundary being traction-free. Let $\mathcal{V}(x)$ denote the portion of the cylinder with abscissa greater than $x$, and let $E(x)$ be the energy of this sub-body:
\begin{equation}
\label{energy}
E(x)=\int_{\mathcal{V}(x)} U \dd[3]{x},
\end{equation}
where $U$ is evaluated at the solution of the variational problem \eqref{energyf1} under constraints \eqref{displ} and \eqref{constraints1}.

Our objective is to establish an upper bound on the energy decay along the cylinder. To this end, we follow the approach of Berdichevsky \cite{berdichevskii1974proof} and introduce an auxiliary problem. For any cross-section $\mathcal{S}(x)$ of the cylinder, we consider a self-balanced traction $\tilde{\tau}_i(x_\alpha ,x)$ applied on this surface.  Solving the corresponding variational problem for the sub-body $\mathcal{V}(x)$ subjected to $\tilde{\tau}_i(x_\alpha ,x)$, we obtain the energy $\tilde{E}(x)$ and the ``surface'' energy $\tilde{\Omega}(x)$:
\begin{equation*}
\tilde{E}(x)=\int_{\mathcal{V}(x)} \tilde{U} \dd[3]{x}, \quad \tilde{\Omega}(x)=\int_{\mathcal{S}(x)} \tilde{U} \dd[2]{x}.
\end{equation*}
We then define a function $\gamma(x)$ as the minimum value of the ratio $\tilde{\Omega}(x)/\tilde{E}(x)$ over all possible self-balanced tractions $\tilde{\tau}_i(x_\alpha ,x)$:
\begin{equation}
\label{senergy}
\gamma (x)=\min_{\tilde{\tau}_i(x_\alpha ,x)} \frac{\tilde{\Omega}(x)}{\tilde{E}(x)}.
\end{equation}
This function $\gamma(x)$, which depends only on the geometry and material properties of the cylinder and has the dimension (length)$^{-1}$, provides a measure of the energy decay rate.  Specifically, we can show that:
\begin{equation}
\label{decay}
E(x)\le E(0) \exp \Bigl[ -\int_0^x \gamma(x)\dd{x} \Bigr] .
\end{equation}
To prove this inequality, we first note that for any $x$,
\begin{equation}
\label{gammaE}
\gamma(x) E(x) \le \Omega (x)=\int_{\mathcal{S}(x)} U \dd[2]{x},
\end{equation}
where $\Omega(x)$ is the ``surface'' energy corresponding to the true traction $\tau_i(x_\alpha ,x)$. This inequality follows directly from the definition of $\gamma(x)$ in Eq.~\eqref{senergy}, as the true traction is just one specific choice among all possible self-balanced tractions.
Next, we differentiate the integral in Eq.~\eqref{energy} with respect to $x$, keeping in mind that the boundary of the integration domain $\mathcal{V}(x)$ depends on $x$. This yields the following identity:
\begin{equation}
\label{dE}
\dv{E}{x}=-\int_{\mathcal{S}(x)} U \dd[2]{x}.
\end{equation}
Combining \eqref{gammaE} and \eqref{dE}, we obtain:
\begin{equation*}
\gamma(x)E(x)+\dv{E}{x}\le 0,
\end{equation*}
This differential inequality can be integrated to yield the desired energy decay estimate in Eq.~\eqref{decay}.

For a semi-infinite functionally graded piezoelectric cylinder with materially uniform cross-sections, $\gamma(x)$ is obviously constant, and the energy decays exponentially:
\begin{equation*}
E(x)\le E(0)\exp (-\gamma x).
\end{equation*}

\section{Lower bound for the energy decay rate}
\subsection{Lower bounds for the energy densities}

To establish a lower bound for the energy decay rate in a functionally graded piezoelectric cylinder, we first derive lower bounds for the internal and complementary energy densities, denoted by $U(x_\alpha,\varepsilon_{ij},D_i)$ and $U^*(x_\alpha ,\sigma_{ij},E_i)$, respectively.

We begin with the internal energy density, $U(x_\alpha,\varepsilon_{ij},D_i)$, which can be expressed as the Young-Fenchel transformation of the negative electric enthalpy, $-W(x_\alpha,\varepsilon_{ij},E_j)$, with respect to the electric field, $E_i$:
\begin{equation}
\label{Legendre}
U(x_\alpha,\varepsilon_{ij},D_i)=\max_{E_i}[D_iE_i+W(x_\alpha,\varepsilon_{ij},E_j)].
\end{equation}
Here, the electric enthalpy density, $W(x_\alpha,\varepsilon_{ij},E_j)$, is given by:
\begin{equation*}
W(x_\alpha,\varepsilon_{ij},E_j)=\frac{1}{2}c^E_{ijkl}\varepsilon_{ij}\varepsilon_{kl}-e_{ijk}E_i\varepsilon_{jk}-\frac{1}{2}\epsilon^S_{ij}E_iE_j.
\end{equation*}
Utilizing the constitutive relation between electric displacement, electric field, and strain, which corresponds to the condition for the maximum in \eqref{Legendre}:
\begin{equation*}
D_i=-\pdv{W}{E_i}=e_{ijk}\varepsilon_{jk}+\epsilon^S_{ij}E_j
\end{equation*}
we obtain the following expression for the internal energy density:
\begin{equation}
\label{Ueq}
U(x_\alpha,\varepsilon_{ij},D_i)=\frac{1}{2}c^E_{ijkl}\varepsilon_{ij}\varepsilon_{kl}+\frac{1}{2}\epsilon^S_{ij}E_iE_j.
\end{equation}
The positive definiteness of the second term in \eqref{Ueq} allows us to establish the lower bound:
\begin{equation}
\label{Uineq}
\frac{1}{2}c^E_{ijkl}\varepsilon_{ij}\varepsilon_{kl} \le U(x_\alpha,\varepsilon_{ij},D_i).
\end{equation}

Since the left-hand side of \eqref{Uineq} is a strictly positive definite quadratic form in $\varepsilon_{ij}$, we can introduce positive constants $A_1,A_2,A_3$ such that
\begin{equation}
\label{Ulb}
\frac{1}{2}(A_1\varepsilon_{\alpha \beta}\varepsilon_{\alpha \beta}+A_2\varepsilon^2+2A_3\varepsilon_\alpha \varepsilon_\alpha)\le U(x_\alpha,\varepsilon_{ij},D_i)
\end{equation}
holds for all $\varepsilon_{ij}$ and $D_i$. To illustrate the determination of these constants, we consider a transversely isotropic functionally graded piezoelectric material, for which the left-hand side of \eqref{Uineq} can be expressed as:
\begin{equation*}
\frac{1}{2}c^E_{ijkl}\varepsilon_{ij}\varepsilon_{kl}=\frac{1}{2}[c_1(\varepsilon_{\alpha \alpha})^2+2c_2 \varepsilon_{\alpha \beta}\varepsilon_{\alpha \beta}+c_3\varepsilon^2+2c_4 \varepsilon \varepsilon_{\alpha \alpha}+4c_5\varepsilon_\alpha \varepsilon_\alpha ],
\end{equation*}
where $c_1,\ldots ,c_5$ are five independent elastic stiffnesses, which are functions of $x_\alpha$:
\begin{equation*}
c_1=c^E_{1122},\, c_2=(c^E_{1111}-c^E_{1122})/2,\, c_3=c^E_{3333},\, c_4=c^E_{1133},\, c_5=c^E_{1313}.
\end{equation*}
Our aim is to find constants $A_1$, $A_2$, $A_3$ that satisfy the inequality: 
\begin{equation*}
\begin{split}
&c_1(\varepsilon_{\alpha \alpha})^2+2c_2 \varepsilon_{\alpha \beta}\varepsilon_{\alpha \beta}+c_3\varepsilon^2+2c_4 \varepsilon \varepsilon_{\alpha \alpha}+4c_5\varepsilon_\alpha \varepsilon_\alpha  
\\
&\ge A_1\varepsilon_{\alpha \beta}\varepsilon_{\alpha \beta}+A_2\varepsilon^2+2A_3\varepsilon_\alpha \varepsilon_\alpha.
\end{split}
\end{equation*}
Recognizing the independence of the strain components $\varepsilon_\alpha$, we can choose $A_3=2\min_{x_\alpha}c_5(x_\alpha)\equiv 2\bar{c}_5$. To isolate independent components within $\varepsilon_{\alpha \beta}$, we express them as $\varepsilon_{\alpha \beta}=\varepsilon^\prime_{\alpha \beta}+\frac{1}{2}\delta_{\alpha \beta}\varepsilon_{\gamma \gamma}$, where $\varepsilon^\prime_{\alpha \beta}$ represents the 2-D strain deviator satisfying $\varepsilon^\prime_{\alpha \alpha}=0$. This leads to the inequality:
\begin{equation*}
\begin{split}
&(c_1+c_2) (\varepsilon_{\alpha \alpha})^2+2c_2\varepsilon^\prime_{\alpha \beta}\varepsilon^\prime_{\alpha \beta}+c_3 \varepsilon^2+2c_4 \varepsilon \varepsilon_{\alpha \alpha}
\\
&\ge A_1\varepsilon^\prime_{\alpha \beta}\varepsilon^\prime_{\alpha \beta}+\frac{A_1}{2}(\varepsilon_{\alpha \alpha})^2+A_2\varepsilon^2.
\end{split}
\end{equation*} 
Due to the independence of $\varepsilon^\prime _{\alpha \beta}$, we can select $A_1\le 2\min_{x_\alpha}c_2(x_\alpha)\equiv 2\bar{c}_2$. Setting $A_1=2\alpha \bar{c}_2$, with $\alpha\in (0,1)$, requires that the coefficient $A_2$ satisfy:
\begin{equation}
\label{A2in}
(c_1+(1-\alpha )\bar{c}_2)(\varepsilon_{\alpha \alpha})^2+2c_4 \varepsilon \varepsilon_{\alpha \alpha}+c_3 \varepsilon^2\ge A_2\varepsilon^2.
\end{equation}
By completing the positive full square with the first two terms on the left-hand side, we find that
\begin{equation*}
A_2=\min_{x_\alpha} \Bigl[ c_3-\frac{c_4^2 }{c_1+(1-\alpha)\bar{c}_2}\Bigr] 
\end{equation*}
satisfies the inequality \eqref{A2in} for all $\varepsilon$ and $\varepsilon_{\alpha \alpha}$, provided $c_1+(1-\alpha )\bar{c}_2\ge 0$. Therefore, the lower bound \eqref{Ulb} holds for all $\varepsilon_{ij}$ and $D_i$ with the constants
\begin{equation*}
A_1=2\alpha \bar{c}_2, \quad A_2=\min_{x_\alpha} \Bigl[ c_3-\frac{c_4^2 }{c_1+(1-\alpha)\bar{c}_2}\Bigr] ,\quad A_3=2\bar{c}_5,
\end{equation*}
where $\alpha\in (0,1)$.

In addition to the internal energy density, we introduce the complementary energy density, $U^*(x_\alpha ,\sigma_{ij},E_i)$, also known as the Gibbs function \cite{le1999vibrations}, and the elastic enthalpy density, $F(x_\alpha ,\sigma_{ij},D_i)$. These are defined through Young-Fenchel transformations of the internal energy density, $U(x_\alpha,\varepsilon_{ij},D_i)$:
\begin{equation*}
\begin{split}
U^*(x_\alpha ,\sigma_{ij},E_i)&=\max_{\varepsilon_{ij},D_i} [\sigma_{ij}\varepsilon_{ij}+E_iD_i-U(x_\alpha,\varepsilon_{ij},D_i)]
\\
&=\frac{1}{2}s^E_{ijkl}\sigma_{ij}\sigma_{kl}+d_{ijk}E_i\sigma_{jk}+\frac{1}{2}\epsilon^T_{ij}E_iE_j,
\\
F(x_\alpha ,\sigma_{ij},D_i)&=-\max_{\varepsilon_{ij}} [\sigma_{ij}\varepsilon_{ij}-U(x_\alpha,\varepsilon_{ij},D_i)]
\\
&=-\frac{1}{2}s^D_{ijkl}\sigma_{ij}\sigma_{kl}+g_{ijk}D_i\sigma_{jk}+\frac{1}{2}\beta^T_{ij}D_iD_j.
\end{split}
\end{equation*}
It is important to note that  $U^*(x_\alpha ,\sigma_{ij},E_i)$ is a strictly positive quadratic form in $\sigma_{ij}$ and $E_i$, while $F(x_\alpha ,\sigma_{ij},D_i)$ is indefinite, satisfying $F(\vb{0},\vb{D})\ge 0$ and $F(\vb*{\sigma},\vb{0})\le 0$. Expressing $U^*(x_\alpha ,\sigma_{ij},E_i)$ as the Young-Felchen transformation of the elastic enthalpy, $F((x_\alpha ,\sigma_{ij},D_i)$, with respect to $D_i$
\begin{equation}
\label{YFF}
U^*(x_\alpha ,\sigma_{ij},E_i)=\max_{D_i} [E_iD_i-F(x_\alpha,\sigma_{ij},D_i)],
\end{equation}
and employing the constitutive equation corresponding to the condition for the maximum in \eqref{YFF}
\begin{equation*}
E_i=\pdv{F}{D_i}=g_{ijk}\sigma_{jk}+\beta^T_{ij}D_j,
\end{equation*}
we arrive at:
\begin{equation*}
U^*(x_\alpha ,\sigma_{ij},E_i)=\frac{1}{2}s^D_{ijkl}\sigma_{ij}\sigma_{kl}+\frac{1}{2}\beta^T_{ij}D_iD_j.
\end{equation*}
The positive definiteness of $\frac{1}{2}\beta^T_{ij}D_iD_j$ implies the following lower bound for $U^*(x_\alpha,\sigma_{ij},E_i)$:
\begin{equation}
\label{Ustineq}
\frac{1}{2}s^D_{ijkl}\sigma_{ij}\sigma_{kl} \le U^*(x_\alpha,\sigma_{ij},E_i).
\end{equation}
Since the left-hand side of \eqref{Ustineq} is a strictly positive definite quadratic form in $\sigma_{ij}$, we can introduce positive constants $B_1,B_2,B_3$ such that
\begin{equation}
\label{Ustlb}
\frac{1}{2}(B_1\sigma_{\alpha \beta}\sigma_{\alpha \beta}+B_2\sigma^2+2B_3\sigma_\alpha \sigma_\alpha)\le U^*(x_\alpha,\sigma_{ij},E_i)
\end{equation}
holds for all $\sigma_{ij}$ and $E_i$. 

For a transversely isotropic functionally graded piezoelectric material, the compliance tensor $s^D_{ijkl}$ can be expressed in terms of five independent elastic compliances, $s_1,\ldots ,s_5$, such that
\begin{equation*}
\frac{1}{2}s^D_{ijkl}\sigma_{ij}\sigma_{kl}=\frac{1}{2}[s_1(\sigma_{\alpha \alpha})^2+2s_2 \sigma_{\alpha \beta}\sigma_{\alpha \beta}+s_3\sigma^2+2s_4 \sigma \sigma_{\alpha \alpha}+4s_5\sigma_\alpha \sigma_\alpha ].
\end{equation*}
These compliances are defined as:
\begin{equation*}
s_1=s^D_{1122},\, s_2=(s^D_{1111}-s^D_{1122})/2,\, s_3=s^D_{3333},\, s_4=s^D_{1133},\, s_5=s^D_{1313}.
\end{equation*}
We seek to find constants $B_1$, $B_2$, and $B_3$ satisfying the inequality:
\begin{equation*}
\begin{split}
&s_1(\sigma_{\alpha \alpha})^2+2s_2 \sigma_{\alpha \beta}\sigma_{\alpha \beta}+s_3\sigma^2+2s_4 \sigma \sigma_{\alpha \alpha}+4s_5\sigma_\alpha \sigma_\alpha   
\\
&\ge B_1\sigma_{\alpha \beta}\sigma_{\alpha \beta}+B_2\sigma^2+2B_3\sigma_\alpha \sigma_\alpha.
\end{split}
\end{equation*}
Following a similar line of reasoning as in the previous case, we can show that
\begin{equation*}
B_1=2\beta \bar{s}_2, \quad B_2=\min_{x_\alpha} \Bigl[ s_3-\frac{s_4^2 }{s_1+(1-\beta)\bar{s}_2}\Bigr],\quad B_3=2\bar{s}_5,
\end{equation*}
with $\beta\in (0,1)$, satisfy the lower bound \eqref{Ustlb} for all $\sigma_{ij}$ and $E_i$, provided $s_1+(1-\beta)\bar{s}_2\ge 0$. Finally, we note the identity 
\begin{equation}
\label{identity}
U^*(x_\alpha,\sigma_{ij}(\varepsilon_{ij},D_i),E_i(\varepsilon_{ij},D_i))=U(x_\alpha,\varepsilon_{ij},D_i),
\end{equation}
which holds when $\sigma_{ij}$ and $E_i$ satisfy the constitutive equations \eqref{constitutive}.

\subsection{Lower bound for the energy decay rate}
Having established lower bounds for the energy densities, we now proceed to derive a lower bound for the energy decay rate, $\gamma (x)$, in a piezoelectric cylinder under self-balanced traction applied at its edge. 

We begin by introducing a positive constant, $b(x)$, associated with the following inequality
\begin{equation}
\label{ineqb}
b(x) \int_{\mathcal{S}(x)} w^2\dd[2]x\le \int_{\mathcal{V}(x)}U\dd[3]x,
\end{equation}
which holds for any domain $\mathcal{V}(x)$ with base $\mathcal{S}(x)$. This inequality stems from the lower bound \eqref{Ulb} and the trace, Poincaré-Wirtinger, and Korn inequalities \eqref{Korn-Poinc}. To ensure the validity of \eqref{ineqb}, we impose constraints on the displacement field to eliminate rigid body motions. Since only the transverse displacement component, $w$, appears in the left-hand side of \eqref{ineqb}, we need only enforce the following conditions:
\begin{equation}
\label{excl}
\int_{\mathcal{V}(x)}(w_{,\alpha }-w_{\alpha ,x})\dd[3]{x}=0, \quad \int_{\mathcal{V}(x)}w\dd[3]{x}=0.
\end{equation}
We denote by $b$ the largest possible constant (the best constant) and by $\mathcal{V}^\prime(x)$ the sub-domain of $\mathcal{V}(x)$ with base $\mathcal{S}(x)$ and a traction-free end, for which the following inequality holds:
\begin{equation}
\label{ineqbc}
b \int_{\mathcal{S}(x)} w^2\dd[2]x\le \int_{\mathcal{V}^\prime (x)}U\dd[3]x.
\end{equation}
Our objective is to determine a lower bound for the energy decay rate, $\gamma(x)$, in terms of this best constant, $b$.  

Consider a self-balanced traction, $\tau_i(x_\alpha,0)$, applied to the edge $\mathcal{S}_0$ of the cylinder. This traction generates displacement and stress fields within the cylinder, with $\tau_i(x_\alpha,x)=\sigma_{ij}n_j$ representing the self-balanced traction acting on the cross-section $\mathcal{S}(x)$. Following the principle that the total energy decreases when a deformed body is joined to an undeformed body along a traction-free surface \cite{berdichevskii1974proof}, we first calculate the energy, $E^\prime (x)$, of the sub-body occupying $\mathcal{V}^\prime(x)$, for which inequality \eqref{ineqbc} is satisfied. Given the quadratic nature of $U(x_\alpha,\varepsilon_{ij},D_i)$ in $\varepsilon_{ij}$ and $D_i$, we have:
\begin{equation}
\label{energyx}
\begin{split}
E^\prime (x)&=\int_{\mathcal{V}^\prime (x)} U \dd[3]{x}=\frac{1}{2}\int_{\mathcal{V}^\prime (x)} \Bigl( \pdv{U}{\varepsilon_{ij}}\varepsilon_{ij}+\pdv{U}{D_i} D_i \Bigr)\dd[3]{x}
\\
&=\frac{1}{2}\int_{\mathcal{V}^\prime (x)} ( \sigma_{ij}\varepsilon_{ij}+E_i D_i )\dd[3]{x}.
\end{split}
\end{equation}
We then transform the integrand in \eqref{energyx} by replacing $\varepsilon_{ij}=w_{(i,j)}$ and $D_i=\epsilon_{ijk}\psi_{k,j}$, and integrate by parts. Using the Dirichlet's conditions \eqref{constraints1}$_2$ and the traction-free boundary conditions on the remaining boundary $\partial \mathcal{V}^\prime(x) \raisebox{0.3ex}{\scalebox{0.8}{$\setminus$}} \mathcal{S}(x)$, we obtain: 
\begin{equation}
\label{energy1}
E^\prime (x)=\frac{1}{2}\int_{\mathcal{S}(x)}\sigma_{ij}w_in_j\dd[2]x=-\frac{1}{2}\int_{\mathcal{S}(x)}(\sigma w+\sigma_\alpha w_\alpha) \dd[2]x.
\end{equation}
The minus sign arises from the orientation of the outward unit normal vector on $\mathcal{S}(x)$, which is $\vb{n}=-\vb{e}_3\equiv -\vb{e}$, where $\vb{e}_1$, $\vb{e}_2$, and $\vb{e}$ are the basis vectors of the Cartesian coordinate system.  By adding a rigid body motion to the displacement field in \eqref{energy1}, we can satisfy the constraints \eqref{excl} without altering the energy or the traction boundary conditions. This allows us to further impose the following conditions:
\begin{equation}
\label{exclpl}
\int_{\mathcal{S}(x)}w_\alpha \dd[2]x=0,\quad \int_{\mathcal{S}(x)}(w_{\alpha ,\beta}-w_{\beta ,\alpha})\dd[2]x=0.
\end{equation}
In addition to \eqref{ineqb}, we employ the inequality
\begin{equation}
\label{ineqs}
\lambda^2\int_{\mathcal{S}}w_\alpha w_\alpha \dd[2]{x}\le \int_{\mathcal{S}}\varepsilon_{\alpha \beta} \varepsilon_{\alpha \beta}\dd[2]{x},
\end{equation}
which holds for 2-D vector fields subject to constraints \eqref{exclpl}. This inequality follows from the 2-D analogs of the Poincaré-Wirtinger and second Korn's inequalities:
\begin{equation*}
\begin{split}
&\int_\mathcal{S} w_\alpha w_\alpha \dd[2]{x} \le p \int_\mathcal{S} w_{\alpha ,\beta} w_{\alpha ,\beta} \dd[2]{x}
\\
&\int_{\mathcal{S}} w_{\alpha ,\beta }w_{\alpha ,\beta }\dd[2]{x}\le k \int_{\mathcal{S}} \varepsilon _{\alpha \beta } \varepsilon _{\alpha \beta } \dd[2]{x},
\end{split}
\end{equation*}
where $p$ and $k$ are the 2-D Poincare-Wirtinger's and Korn's constants, respectively and $\lambda^2=(pk)^{-1}$.

Applying the Cauchy-Schwarz inequality to the right-hand side of \eqref{energy1}, we obtain:
\begin{equation}
\label{ineqe}
\begin{split}
2E^\prime (x)&\le \Bigl( \int_{\mathcal{S}(x)} \sigma^2\dd[2]{x}\Bigr)^{1/2} \Bigl( \int_{\mathcal{S}(x)} w^2\dd[2]{x}\Bigr)^{1/2}
\\
&+\Bigl( \int_{\mathcal{S}(x)} \sigma_\alpha \sigma_\alpha \dd[2]{x}\Bigr)^{1/2} \Bigl( \int_{\mathcal{S}(x)} w_\alpha w_\alpha \dd[2]{x}\Bigr)^{1/2}.
\end{split}
\end{equation}
By utilizing the lower bounds \eqref{Ustlb} and the identity \eqref{identity}, together with inequalities \eqref{ineqb} and \eqref{ineqs}, we can estimate the integrals in \eqref{ineqe} as follows
\begin{equation*}
\begin{split}
2E^\prime (x)&\le  \Bigl( 2B_2^{-1} \int_{\mathcal{S}(x)} U\dd[2]{x}\Bigr)^{1/2} (b^{-1}E^\prime (x))^{1/2}
\\
&+\Bigl( B_3^{-1} \int_{\mathcal{S}(x)} U\dd[2]{x}\Bigr)^{1/2} \Bigl( \lambda^{-2} \int_{\mathcal{S}(x)} 2A_1^{-1} U\dd[2]{x}\Bigr)^{1/2}
\\
&\le \epsilon E^\prime (x)+ \Bigl( \frac{1}{2}\epsilon ^{-1}B_2^{-1}b^{-1}+(2B_3^{-1}A_1^{-1}\lambda^{-2})^{1/2}\Bigr)\int_{\mathcal{S}(x)} U\dd[2]{x},
\end{split}
\end{equation*}
where $\epsilon$ is a positive number. This leads to the inequality:
\begin{equation}
\label{ineqe1}
E(x)\le E^\prime (x) \le (2-\epsilon)^{-1}\Bigl( \frac{1}{2}\epsilon ^{-1}B_2^{-1}b^{-1}+(2B_3^{-1}A_1^{-1}\lambda^{-2})^{1/2}\Bigr) \int_{\mathcal{S}(x)} U\dd[2]{x}.
\end{equation}
Minimizing the factor on the right-hand side of \eqref{ineqe1} with respect to $\epsilon$ and substituting the result into the expression for the energy decay rate, we obtain the lower bound:
\begin{equation}
\label{lbg}
8bB_2 \Bigl[ 1+\Bigl( 1+4bB_2\Bigl( \frac{2pk}{A_1B_3}\Bigr) ^{1/2}\Bigr)^{1/2}\Bigr]^{-2}\le \gamma (x).
\end{equation}

Since the expression in the left-hand side of \eqref{lbg} is the increasing function of $B_2$, $A_1$, and $B_3$, we maximize this lower bound by selecting the largest possible values for these constants, which are achieved at $\alpha=1$ and $\beta=0$. For the transversely isotropic functionally graded piezoelectric materials, this yields
\begin{equation*}
A_1=2\bar{c}_2,\quad B_2=\min_{x_\alpha} \Bigl[ s_3-\frac{s_4^2 }{s_1+\bar{s}_2}\Bigr],\quad B_3=2\bar{s}_5.
\end{equation*}
Consequently, the lower bound \eqref{lbg} becomes:
\begin{equation}
\label{lbg1}
8bB_2 \Bigl[ 1+\Bigl( 1+4bB_2\Bigl( \frac{2pk}{A_1B_3}\Bigr) ^{1/2}\Bigr)^{1/2}\Bigr]^{-2}\le \gamma (x).
\end{equation} 

Berdichevsky \cite{berdichevskii1974proof} derived a lower bound for $b(x)$, which, after correcting for misprints\footnote{In \cite{berdichevskii1974proof} there are misprints in Eqs.~(6.15) and (6.17): The factor $B_1$ must be replaced by $A_1$.}, takes the form
\begin{equation}
\label{lbb}
b(x) \ge \frac{1}{2}\Bigl( \frac{12(5A_3+A_1h^2\lambda^2)}{5A_1A_3h^3 \lambda^2 p^{-1}(1-\kappa)}+\frac{13h}{35A_2}\Bigr) ^{-1},
\end{equation}
where $h$ is the length of the sub-domain $\mathcal{V}^\prime (x)$, and $\kappa$ is the best constant in the inequality
\begin{equation*}
\Bigl( \int_\mathcal{S} e_{\alpha \beta}u_{,\alpha }x_\beta \dd[2]{x} \Bigr)^2 \le \kappa \int_\mathcal{S} u_{,\alpha }u_{,\alpha } \dd[2]{x} \int_\mathcal{S} x_{\alpha }x_{\alpha } \dd[2]{x},
\end{equation*}
with $e_{\alpha \beta}$ denoting the 2-D permutation symbol. For a circular cross-section, $\kappa =0$. The derivation of the lower bound \eqref{lbb} relies on certain inequalities and the result from \cite{berdichevskii1973energy} that the elastic energy of the Reissner model provides a lower bound for the elastic energy of a cylindrical body. By maximizing the right-hand side of \eqref{lbb} with respect to $h$, we can determine the best constant $b$, provided the 2-D Poincaré-Wirtinger and Korn constants are known. 

For semi-infinite cylinders with a circular cross-section, we have $\kappa =0$, and the 2-D Poincaré-Wirtinger and Korn constants are given by \cite{payne1961korn}
\begin{equation*}
p=\frac{r^2}{j^2}, \quad k=4,
\end{equation*}
where $j=1.845$ is the first zero of the derivative of the Bessel function $J_1(x)$, and $r$ is the radius of the cross-section. Consequently, $\lambda ^2=(pk)^{-1}=j^2/(4r^2)$. Substituting these values into inequality \eqref{lbb}, we obtain:
\begin{equation}
\label{lbb1}
b(x) \ge \frac{A_2}{2r}\Bigl( \frac{12(5a_3+a_1\zeta^2j^2/k)}{5a_1a_3\zeta^3 j^4/k}+\frac{13\zeta }{35}\Bigr) ^{-1},
\end{equation}
where
\begin{equation*}
a_1=\frac{A_1}{A_2},\quad a_3=\frac{A_3}{A_2}, \quad \zeta =\frac{h}{r}.
\end{equation*}

\begin{table}[h]
\centering
\begin{tabular}{|c|c|c|c|c|c|c|c|c|c|}
\hline
$s^E_{11}$ & $s^E_{13}$ & $s^E_{33}$ & $s^E_{55}$ & $s^E_{66}$ & $s^D_{11}$ & $s^D_{13}$ & $s^D_{33}$ & $s^D_{55}$ & $s^D_{66}$ \\
\hline
12.3 & -5.31 & 15.5 & 39 & 32.7 & 10.9 & -2.1 & 7.9 & 19.3 & 32.7 \\
\hline
\end{tabular}
\caption{Elastic compliances $s^E_{ij}$ and $s^D_{ij}$ of PZT-4.}
\label{table:1}
\end{table}

To illustrate the calculation of the lower bound for the energy decay rate, we consider a homogeneous PZT-4 piezoceramic cylinder polarized in the $x_3$-direction. The elastic compliances of PZT-4 (in units of $10^{-12}$m$^2/$N) are provided in Table~\ref{table:1}, using Voigt's shorthand index notation \cite{berlincourt1964piezoelectric}. The corresponding elastic stiffnesses, $c^E_{ij}$, are computed as follows:
\begin{equation*}
\begin{pmatrix}
  c^E_{11}  &  c^E_{12}  &  c^E_{13} \\
  c^E_{12}  &  c^E_{11}  &  c^E_{13} \\
  c^E_{13}  &  c^E_{13}  &  c^E_{33}   
  \end{pmatrix} = 
\begin{pmatrix}
  s^E_{11}  &  s^E_{12}  &  s^E_{13} \\
  s^E_{12}  &  s^E_{11}  &  s^E_{13} \\
  s^E_{13}  &  s^E_{13}  &  s^E_{33}   
  \end{pmatrix}^{-1},
\end{equation*}
where $s^E_{12}=s^E_{11}-s^E_{66}/2$, and
\begin{equation*}
c^E_{55}=\frac{1}{s^E_{55}},\quad c^E_{66}=\frac{1}{s^E_{66}}.
\end{equation*}
The resulting elastic stiffnesses, $c^E_{ij}$, (in units of $10^{10}$N/m$^2$) are:
\begin{equation*}
c^E_{11}=13.9,\quad c^E_{12}=7.78,\quad c^E_{13}=7.43,\quad c^E_{33}=11.54,\quad c^E_{55}=2.56,\quad c^E_{66}=3.06.
\end{equation*}
Using these values, we determine the constants $A_1,A_2,A_3,B_2,B_3$: 
\begin{equation*}
\begin{split}
&A_1=2c_2=2c^E_{66}=6.12\times 10^{10}\text{N/m}^2
\\
&A_2=c_3-\frac{c_4^2}{c_1}=c^E_{33}-\frac{(c^E_{13})^2}{c^E_{12}}=4.45\times 10^{10}\text{N/m}^2
\\
&A_3=2c_5=2c^E_{55}=5.13 \times 10^{10}\text{N/m}^2,
\\
&B_2=s_3-\frac{s_4^2}{s_1+s_2}=s^D_{33}-\frac{(s^D_{13})^2}{s^D_{12}+s^D_{66}/4}=6.2817 \times 10^{-12}\text{m}^2/\text{N},
\\
&B_3=2s_5=\frac{1}{2}s^D_{55}=9.65 \times 10^{-12}\text{m}^2/\text{N}.
\end{split}
\end{equation*}
From the calculated values, we obtain:
\begin{equation*}
a_1=1.3738,\quad a_3=1.1519.
\end{equation*}
Using these numerical values, we determine the best constant $b$ by maximizing the right-hand side of inequality \eqref{lbb1}. For the PZT-4 material, this yields $b=0.3664 A_2/r$, with the maximum achieved at $h/r=2.414$. Substituting this value of $b$ into the lower bound \eqref{lbg1}, we arrive at:
\begin{equation*}
0.1016\, r^{-1}\le \gamma .
\end{equation*}
This illustrates how the lower bound for the energy decay rate can be explicitly computed for a specific material.  The same methodology, based on inequalities \eqref{lbb1} and \eqref{lbg1}, can be applied to any functionally graded piezoelectric material, provided the coefficients  $A_1,A_2,A_3,B_2,B_3$ are known.

\section{Conclusion and remarks}
This study established a lower bound for the energy decay rate in a piezoelectric cylinder of circular cross-section. This bound, proportional to the inverse of the radius, ensures significant energy decay over distances comparable to the radius. This result is relevant for error estimation in piezoelectric rod theories \cite{le1999vibrations,berdichevsky2009variational} and provides an estimate for the stress decay rate, which is half that of the energy. The methodology can be extended to other geometries, such as cone-shaped bodies, where a power-law decay is expected \cite{berdichevskii1974proof}. Future work will focus on determining Poincaré-Wirtinger and Korn constants for various cross-sections to enable broader application of these findings.

\end{document}